\documentclass[5p, twocolumn,sort&compress]{elsarticle}

\usepackage{hyperref,amsmath,amssymb}

\journal{Physics Letters B}

\makeatletter

\def\mypreprint#1{\gdef\@mypreprint{#1}}
\let\@mypreprint\@empty

\def\@author#1{\g@addto@macro\elsauthors{\Large%
    \def\baselinestretch{1}%
    \upshape\authorsep#1\unskip\textsuperscript{%
      \ifx\@fnmark\@empty\else\unskip\sep\@fnmark\let\sep=,\fi
      \ifx\@corref\@empty\else\unskip\sep\@corref\let\sep=,\fi
      }%
    \def\authorsep{\unskip,\space}%
    \global\let\@fnmark\@empty
    \global\let\@corref\@empty
    \global\let\sep\@empty}%
    \@eadauthor={#1}
}

\long\def\MaketitleBox{%
  \resetTitleCounters
  \def\baselinestretch{1}%
  \begin{flushright}%
  \small\@mypreprint%
  \end{flushright}%
  \begin{center}%
   \def\baselinestretch{1}%
    \LARGE\@title\par\vskip18pt
    \normalsize\elsauthors\par\vskip10pt
    \footnotesize\itshape\elsaddress\par\vskip36pt
    \hrule\vskip12pt
    \ifvoid\absbox\else\unvbox\absbox\par\vskip10pt\fi
    \ifvoid\keybox\else\unvbox\keybox\par\vskip10pt\fi
    \hrule\vskip12pt
    \end{center}%
  }

\long\def\@address#1{\g@addto@macro\elsauthors{%
    \def\baselinestretch{1}%
    \addsep\addsep\normalsize\itshape#1\def\addsep{\par\vskip6pt}%
    \def\authorsep{\par\vskip8pt}}}

\def\ps@pprintTitle{%
 \let\@oddhead\@empty
 \let\@evenhead\@empty
 \def\@oddfoot{}%
 \let\@evenfoot\@oddfoot}
 
\newbox\absbox
\renewenvironment{abstract}{\global\setbox\absbox=\vbox\bgroup
  \hsize=\textwidth\def\baselinestretch{1}%
  \noindent\unskip\textsc{\large Abstract}
 \par\medskip\noindent\unskip\ignorespaces}
 {\egroup}

\renewcommand\appendix{\par
  \setcounter{section}{0}%
  \setcounter{subsection}{0}%
  \setcounter{equation}{0}
  \gdef\thefigure{\arabic{figure}}%
  \gdef\thetable{\@Alph\c@section.\arabic{table}}%
  \gdef\thesection{\appendixname~\@Alph\c@section}%
  \@addtoreset{equation}{section}%
  \gdef\theequation{\@Alph\c@section.\arabic{equation}}%
  \addtocontents{toc}{\string\let\string\numberline\string\tmptocnumberline}{}{}
}

\def\appendixname{Appendix}

\gdef\emailauthor#1#2{\stepcounter{ead}%
     \g@addto@macro\@elseads{\raggedright%
      \let\corref\@gobble
      \eadsep\texttt{#1}\def\eadsep{\unskip,\space}}%
}

\renewcommand\@makefntext[1]{#1}

\def\footnoterule{\kern-3\p@
  \hrule \@width 2in \kern 6\p@} 

\makeatother

\newdefinition{cwdef}{Remark}

\begin{document}

\begin{frontmatter}

\title{Clockwork without supersymmetry}

\author{Daniele Teresi}
\ead{Daniele.Teresi@ulb.ac.be}
\address{Service de Physique Th\'eorique, Universit\'e Libre de Bruxelles,\\
Boulevard du Triomphe, CP225, 1050 Brussels, Belgium} 

\mypreprint{ULB-TH/18-01}

\begin{abstract}
The clockwork mechanism would be completely spoiled by the presence of a cosmological constant in the bulk of its 5D construction, or by analogous terms in the discrete case. It is believed that supersymmetry is required to forbid the appearance of these fatal terms, thus apparently tying the fate of the clockwork to the presence of supersymmetry. In this letter we argue that a robust clockwork can be obtained also without supersymmetry, by providing a model for the clockwork/linear dilaton EFT where this protection is  instead related to diffeomorphism invariance in a higher number of extra dimensions. We show that the clockwork mechanism is a quite general setup that emerges, rather minimally, from pure gravity in the presence of $D-5$ additional sufficiently flat extra dimensions. The linear dilaton model is obtained asymptotically in the limit of infinitely many extra dimensions. We then study the finite-$D$ theory, which is a deformation of the linear dilaton one, but still a clockwork theory for $D$ large enough.
\end{abstract}

\end{frontmatter}

\makeatletter
\renewcommand\@makefntext[1]{\leftskip=0em\hskip1em\@makefnmark\space #1}
\makeatother

\section{Introduction and fragility of the clockwork}\label{sec:intro}
The clockwork mechanism~\cite{Choi:2015fiu,Kaplan:2015fuy,Giudice:2016yja} is an elegant way to generate hierarchies $\sim X$ in a theory with no hierarchy in the fundamental parameters, by making use of a chain-like structure of interactions between a number $\sim \log X$ of fields. Particularly interesting is the limit in which these different fields are identified with a \emph{single} field in different points of an extra dimension. The discrete model is recovered when this fifth compactified extra dimension is deconstructed~\cite{ArkaniHamed:2001ca}, whereas for a continuous extra dimension one can also incorporate  gravity consistently and address the hierarchy problem~\cite{Giudice:2016yja}. Recently, a number of theoretical developments have appeared~\cite{Hambye:2016qkf,Craig:2017cda,Giudice:2017suc,Kehagias:2017grx,Antoniadis:2017wyh,Choi:2017ncj,Giudice:2017fmj} together with phenomenological applications, among which to dark matter~\cite{Hambye:2016qkf,Teresi:2017yrp,Kim:2017mtc}, axion~\cite{Giudice:2016yja,Farina:2016tgd,Coy:2017yex} and neutrino physics~\cite{Hambye:2016qkf,Park:2017yrn,Carena:2017qhd,Ibarra:2017tju}, collider searches~\cite{Giudice:2017fmj} and other phenomenological situations~\cite{Kehagias:2016kzt,Ahmed:2016viu,vonGersdorff:2017iym,Ben-Dayan:2017rvr,Im:2017eju,Lee:2017fin,Craig:2017ppp,Patel:2017pct}.

In the extreme, but simplifying, case in which the parameters along the chain/extra dimension $y$ are exactly equal to each other, one obtains that the clockwork mechanism is due, in the Einstein frame, to the curved metric:
\begin{equation}
ds^2 = e^{\frac{4 k y}{3}} (d x ^2 + dy^2) \;,
\end{equation}
whereas, in the Jordan frame, to a flat Minkowski metric and the interaction with the linear dilaton background $\propto e^{S}$, with $S = 2 k y$. This setup originates from the linear dilaton model, in which one extra dimension is compactified on an orbifold, $0 \leq y \leq \pi R$, so that spacetime has the structure 
\begin{equation}\label{eq:5D_structure}
\mathcal{M}_{CW} = \mathbb R^4 \times \mathcal{S}_1/\mathbb{Z}_2 \;,
\end{equation}
with two 4-branes at the orbifold fixed points $y = 0, \pi R$ and a dilaton $S$ in the bulk, interacting with a costant vacuum energy density $\rho_V = - 2 M_5^3 k^2$, with Jordan-frame action~\cite{Giudice:2016yja}:
\begin{equation}\label{eq:LD}
\mathcal{S} = \frac{M_5^3}{2}  \int_{\mathcal{M}_{CW}} \hspace{-1.5em} d^5 x \, \sqrt{-g} \, e^S \big( \mathcal{R} \,+\, \partial_m S \, \partial^m S \,+\, 4 k^2 \big) \;.
\end{equation}
The effective 4D Planck mass is $M_P = e^{2 \pi k R} M_5^{3/2} /k^{1/2}$, so that the hierarchy problem is solved for $k R \approx 10$. It was later shown~\cite{Hambye:2016qkf} (see also~\cite{Craig:2017cda}) that the same clockwork phenomenon occurs in a flat extra dimension in the presence of a bulk mass for the ``clockworked'' field. However,  the two constructions turn out to be equivalent~\cite{Giudice:2017suc}, at least in the EFT sense, being related by a field redefinition. More precisely, from the low-energy/EFT perspective, the consistent definition of the clockwork is:
 \emph{a theory with no exponential hierarchy in the fundamental parameters along the chain/extra dimension, that gives rise to an exponential hierarchy between the coupling of a symmetry-protected light mode and of heavier states to the same external sector}, for instance the Standard Model. This is indeed invariant under field redefinitions, and independent on UV features beyond the EFT approach.

The linear dilaton model has a rather peculiar Lagrangian~\eqref{eq:LD} and thus it is interesting to investigate its possible origins in the UV. It is known that this model can come from   string-theoretical considerations. For instance, it can be motivated by string theory in a non-critical number of dimensions (see e.g.~\cite{Giudice:2017suc}) or, by the holographic dual of critical string theory in its Little String Theory limit, i.e. with string coupling tending to zero~\cite{Antoniadis:2001sw}. One may wonder if there exist simple non-stringy UV origins of the clockwork/linear-dilaton EFT. In this letter we will answer positively to this question.

The continuum clockwork mechanism requires, in addition to the bulk Lagrangian~\eqref{eq:LD}, the presence of tuned brane terms:
\begin{equation}
\mathcal{S}_\partial = M_5^3 \int_{\mathcal{M}_{CW}} \hspace{-1.5em} d^5 x \, \sqrt{-g} \, e^S\frac{4 k}{\sqrt{g_{55}}} \, \Big[ \delta(y) \,-\, \delta(y-\pi R)\Big] \;.
\end{equation} 
These brane terms are required to have a 4D Minkowski spacetime and their tuning is nothing but the usual tuning of the 4D cosmological constant, the same way as in Randall-Sundrum models~\cite{Randall:1999ee}. However, in~\cite{Giudice:2017fmj} it was pointed out that an \emph{additional} serious tuning is implicit: a cosmological constant $\Lambda$ in the 5D bulk would be in principle allowed (and generated radiatively) unless forbidden by symmetries. This would modify drastically the clockwork solution~\footnote{A way to understand intuitively why this happens is the following. In the Einstein frame, the action~\ref{eq:LD} becomes, in the presence of a cosmological constant $\Lambda$,
$$
\mathcal{S} = \frac{M_5^3}{2}  \int_{\mathcal{M}_{CW}} \hspace{-1.5em} d^5 x \, \sqrt{-g} \, \big( \mathcal{R} \,- \frac{1}{3}\, \partial_m S \, \partial^m S \,+\, 4 k^2 \, e^{-\frac{2 S}{3}} -2 \Lambda\big) \;.
$$ Thus, the cosmological constant will dominate the potential when the dilaton term becomes exponentially small, and the linear dilaton solution is lost. To prevent this, one needs $2 |\Lambda| \lesssim 4 k^2 e^{- \frac{4 \pi k R}{3}}$, i.e.~to tune the cosmological constant to an exponentially small value (compare this estimate with~\eqref{eq:tuning}), since $0 \leq S \leq 2 k \pi R$.
}; for instance, the size of the extra dimension $R_\Lambda$ corresponding to the same value of $M_P$ for fixed $M_5, \,k$, would become
\begin{equation}\label{eq:tuning}
\frac{R_\Lambda}{R} \simeq 1 + \frac{27\, \Lambda/k^2}{200 \pi \, k R} \, e^{\frac{4}{3} \pi k R} \;,
\end{equation}
for $\Lambda \ll k^2$. Hence, the generation of an \emph{exponentially} large Planck mass would be spoiled, unless $|\Lambda|/k^2 \lesssim 10^{-16}$. The clockwork mechanism is fragile under the presence of a bulk cosmological constant, and the hierarchy problem seems to have been just shifted into an equivalent tuning for $\Lambda$. For these reasons, in~\cite{Giudice:2017fmj} it was argued that the clockwork solution requires the presence of supersymmetry in the bulk, in order to forbid the appearance of a bulk cosmological constant $\Lambda$, or analogous terms in its discrete version.

While this is certainly a perfectly viable possibility, it seems to imply that the fate of the clockwork mechanism is apparently tied to the presence of supersymmetry. In this letter, instead, we argue that one can have a \emph{robust} clockwork mechanism even without supersymmetry. The would-be-fatal cosmological constant $\Lambda$ is forbidden by diffeomorphism invariance in a number of dimensions $D>5$. Remarkably, for flat additional $D-5$ extra dimensions, pure gravity provides a dilaton field, with action of the same form as~\eqref{eq:LD}, thus providing at the same time a simple, non-stringy, UV origin for the clockwork Lagrangian.

\section{Robust clockwork from pure gravity}
Let us consider pure gravity in a $D$-dimensional spacetime in which, in addition to 5 dimensions with structure~\eqref{eq:5D_structure}, there are additional $D-5$ compact flat dimensions, i.e. \begin{equation}\label{eq:D_structure}
\mathcal{M}_{D} = \mathbb R^4 \times \mathcal{S}_1/\mathbb{Z}_2 \times \mathcal{F}_{D-5}\;.
\end{equation}
For definiteness, for now let us take the compact $D-5$ manifold to be a hypercubic flat torus $\mathcal{F}_{D-5} = \mathcal{S}_1^{D-5}$, where each of these extra dimensions have length $L$. Later we will generalize this. Thus, we take the background  metric to have the form
\begin{equation}
\widetilde{g}_{MN} = \begin{pmatrix}
g_{mn} & & \\
 & e^{2 \tau} & \\
 & & \ddots & \\
 & & & e^{2 \tau}
\end{pmatrix} \;,
\end{equation} 
where the tilde denotes $D$-dimensional objects, and $m,n$ are 5D indices. For $L \ll R$ one can dimensionally reduce the theory from $D$ to 5 dimensions, by considering only the 0-modes in $\mathcal F_{D-5}$, so that all background fields depend only on the fifth coordinate $y$ (also by virtue of 4D Poincar\'e invariance):
\begin{equation}\label{eq:metric}
d s^2 =g_{m n}(y) \, dx^m dx^n \, + \, e^{2 \tau(y)} \delta_{a b} \, dz^a d z^b \;,
\end{equation}
where $a,b$ are indices in $\mathcal F_{D-5}$, $0 \leq z^a < L$. The consistency of this Kaluza-Klein ansatz is discussed in~\ref{app:KK_consistency}. It is easy to find the Ricci scalar and the determinant for the metric~\eqref{eq:metric}. The former provides the 5D Ricci scalar and kinetic terms for the dilaton field $S \equiv (D-5) \tau$:
\begin{equation}\label{eq:Rtilde}
\widetilde{\mathcal R} = \mathcal R - 2 \, \nabla^2 S -  \frac{D-4}{D-5} \, \partial^m S \, \partial_m S \;.
\end{equation}
The latter provides the dilatonic exponential factor:
\begin{equation}\label{eq:gtilde}
\sqrt{- \widetilde g} = \sqrt{-  g} \, e^S \;.
\end{equation}
This is the key point: a bulk cosmological constant in the spacetime~\eqref{eq:D_structure}, i.e. a term $\sqrt{- \widetilde g} \, \widetilde \Lambda$ in the D-dimensional action, does not generate a cosmological constant in the $5D$ action, but rather  the dilatonic vacuum energy appearing precisely in~\eqref{eq:LD}. Reversing the argument, diffeomorphism invariance in $D$ dimensions (unbroken in the absence of 5D branes) forbids the appearance of a cosmological constant in the $5D$ theory, thus making the (so far potential) clockwork mechanism automatically robust\footnote{One may think that at scales below $\sim 1/L$ fields are essentially 5-dimensional and therefore generate radiatively a 5D cosmological constant. The apparent paradox disappears when one notices that $D$-dimensional diffeomorfism invariance is maintained only when the whole Kaluza-Klein towers are taken into account. While the 0-modes could contribute to a 5D cosmological constant, the overall contribution of the Kaluza-Klein towers must vanish.}. Radiative corrections to $\widetilde{\Lambda}$ modify (radiatively) the clockwork parameter $k$ in~\eqref{eq:LD}, and therefore the clockwork suppression altogether, in a controlled way.

More concretely, we may consider pure gravity with a bulk cosmological constant in the D-dimensional spacetime, and dimensionally reduce it to 5D:
\begin{align}\label{eq:bulk_Lagr}
\mathcal S = & \, \frac{\widetilde{M}_D^{D-2}}{2} \int_{\mathcal{M}_{D}} \hspace{-1.em} d^D X \, \sqrt{- \widetilde g} \, \big( \widetilde{\mathcal{R}} - 2 \widetilde{\Lambda} \big) \\
\longrightarrow & \, \frac{M_5^3}{2}  \int_{\mathcal{M}_{CW}} \hspace{-1.5em} d^5 x \, \sqrt{-g} \, e^S \Big( \mathcal{R} \,+\, \frac{D-6}{D-5} \, \partial_m S \, \partial^m S \,+\, 4 k^2 \Big) \;, \notag
\end{align}
with
\begin{equation}\label{eq:matching}
k^2 \equiv - \frac{\widetilde \Lambda}{2} > 0 \qquad \text{and} \qquad M_5^3 \equiv \widetilde{M}_D^{D-2} L^{D-5} \;.
\end{equation}
The Jordan-frame linear dilaton Lagrangian~\eqref{eq:LD} is obtained, asymptotically, in the limit $D \to \infty$. Going to the Einstein frame, in this limit the bulk equations of motion have the clockwork solution:
\begin{equation}
g_{mn} = e^{\frac{4}{3} k y} \eta_{m n} \;, \qquad S = 2 k y \;,
\end{equation}
protected by the higher-dimensional diffeomorphism invariance. In the Jordan frame, instead, the solution takes on the form 
\begin{equation}\label{eq:sol_JF}
g_{mn} = \eta_{m n} \;, \qquad S = 2 k y \;.
\end{equation}

A few comments are in order. First, in addition to the bulk Lagrangian~\eqref{eq:bulk_Lagr}, one needs to generate also brane terms required to tune to 4D cosmological constant to zero, as mentioned in Sec.~\ref{sec:intro}. These terms are indeed allowed by the setup described above and obtained by considering brane vacuum energies in the D-dimensional theory, dimensionally reduced to 5D:
\begin{align}
&\mathcal S_\partial = \widetilde{M}_D^{D-3} \! \int_{\mathcal{M}_{D}} \hspace{-1.em} d^D X  \, \frac{\sqrt{- \widetilde{g}}}{\sqrt{g_{55}}} \, \Big[ - \widetilde{\Lambda}_0 \, \delta(y) - \widetilde{\Lambda}_{\pi} \, \delta(y- \pi R)\Big] \notag \\
& \longrightarrow  M_5^{2} \! \int_{\mathcal{M}_{CW}} \hspace{-1.em} d^5 x \, \frac{ \sqrt{- g} }{\sqrt{g_{55}}}\, e^S \, \Big[ - \Lambda_0 \, \delta(y) - \Lambda_{\pi} \, \delta(y- \pi R)\Big] \;,
\end{align}
with $\Lambda_{0,\pi} \equiv \widetilde{\Lambda}_{0,\pi} M_5/ \widetilde{M}_D$.  

Second, the compact manifold $\mathcal{F}_{D-5}$ does not need to be a flat torus. Indeed, the formulas~\eqref{eq:Rtilde}, \eqref{eq:gtilde} and~\eqref{eq:bulk_Lagr} (as well as the results in~\ref{app:KK_consistency}) are all valid (or trivially modified) for a more general warped metric 
\begin{equation}\label{eq:metric_general}
d s^2 = g_{m n}(y) \, dx^m dx^n \, + \, e^{2 \tau(y)} \gamma_{a b}(z) \, dz^a d z^b \;,
\end{equation}
as long as the compact manifold $\mathcal{F}_{D-5}$ is Ricci-flat, i.e.~$\mathcal{R}(\gamma_{cd})_{ab}=0$. The matching relation between the Planck masses in~\eqref{eq:matching} is changed into  $M_5^3 \equiv \widetilde{M}_D^{D-2} \mathcal{V}_{D-5}$, with $\mathcal{V}_{D-5}$ being the volume of $\mathcal{F}_{D-5}$. Clearly, the metric $\gamma_{ab}$ contains many moduli fields that determine the geometry and size of $\mathcal{F}_{D-5}$. As in string-theoretic constructions, they need to be stabilized by some (potentially quantum-gravitational) mechanism at a size $\sim L \ll R$. The quantum-gravity expectation $L \sim \widetilde{M}_D^{-1}$ would imply $M_5 \sim \widetilde{M}_D$.

Finally, one may wonder about what happens if the $D-5$ extra dimensions are not flat. As shown in~\ref{app:curved}, in this case the Lagrangian~\eqref{eq:bulk_Lagr} gets deformed. By considering the case of a $(D-5)$-sphere we show that in the limit $D \to \infty$ the curvature radius must go to infinity too. Thus, this seems to indicate that the clockwork occurs if the $D-5$ dimensional manifold is sufficiently flat. Importantly, in~\ref{app:curved} we also argue that the construction considered in this letter does not lead to a tuning analogous to the one of the cosmological constant, because the effect of the deformation given by the curvature does not get exponentially enhanced as in~\eqref{eq:tuning}.

\section{Finite-$D$ quasi-clockwork}
The action~\eqref{eq:bulk_Lagr} reduces to the linear-dilaton one in the limit $D \to \infty$. However, as we are going to show now, for large but finite $D$, this 5D ``quasi-clockwork'' effective field theory, obtained from pure gravity by dimensionally reducing the small $D-5$  dimensions, is anyway a clockwork theory, according to the definition outlined in Sec.~\ref{sec:intro}.

To do so, let us consider the finite-$D$ equations of motion for the action~\eqref{eq:bulk_Lagr}. By virtue of the consistency of the Kaluza-Klein ansatz (see~\ref{app:KK_consistency}) these can be equivalently obtained as the $D$-dimensional Einstein field equations for the metric~\eqref{eq:metric} with a cosmological constant $\widetilde{\Lambda} = - 2 k^2$. By making use of the ansatz:
\begin{equation}
g_{mn} = \begin{pmatrix}
e^{2 a y} \eta_{\mu \nu} & \\
& 1
\end{pmatrix}\;, \qquad \tau = a y\;,
\end{equation}
we find that all the $\mu \nu$, 55, and $ab$ components of the Einstein field equations are satisfied for $a= 2 k / \mathcal{D}$, with $\mathcal{D} \equiv \sqrt{(D-1)(D-2)}$, i.e.
\begin{equation}\label{eq:sol_quasiCW}
g_{\mu \nu} = e^{\frac{4 k y}{\mathcal{D}}}  \eta_{\mu \nu} \;, \quad g_{55}= 1 \;, \quad S= \frac{D-5}{\mathcal{D}} \, 2 k y \;.
\end{equation}
For large $D$ this is a deformation of the Jordan-frame linear-dilaton solution~\eqref{eq:sol_JF}. Notice that, while recovering the linear-dilaton effective field theory as in~\eqref{eq:bulk_Lagr} depends on the assumed hierarchy $L \ll R$, the classical solution~\eqref{eq:sol_quasiCW}, which exhibits the characteristic clockwork suppression, is independent on the value of $L/R$~\footnote{Therefore, even at the (perturbative) quantum level, quantum corrections would be weighted by the exponential dilaton factor, thus not destabilizing the clockwork mechanism.}.

Let us consider a ``clockworked'' scalar field $\phi$. We may start from the $D$-dimensional theory with action
\begin{equation}
\mathcal{S} = - \frac{1}{2} \int_{\mathcal{M}_{D}} \hspace{-1.em} d^D X \, \sqrt{- \widetilde g} \; \widetilde{g}^{MN} \,  \partial_M \widetilde{\phi} \, \partial_N \widetilde{\phi} \;,
\end{equation}
which, after dimensional reduction and in the background~\eqref{eq:sol_quasiCW}, gives the 5D action
\begin{equation}
\mathcal{S} = - \frac{1}{2} \int_{\mathcal{M}_{CW}} \hspace{-1.em} d^5 x \, e^{\frac{D-1}{\mathcal{D}} 2 k y} \, \bigg( e^{-\frac{4 k y}{\mathcal{D}}} \eta^{\mu \nu} \partial_\mu \phi \, \partial_\nu \phi + \partial_y \phi \, \partial_y \phi \bigg) \;.
\end{equation}
The Kaluza-Klein modes are obtained, as usual, by means of the decomposition
\begin{equation}
\phi = \frac{1}{\sqrt{\pi R}} \sum_n \phi_n(x) \psi_n(y)\;, \quad \text{with} \quad \partial_x^2 \phi_n = M_n^2 \phi_n \;.
\end{equation}
Their equation of motion is:
\begin{equation}\label{eq:EoM}
\bigg( \partial_y \,+ \, 2 k \frac{D-1}{\mathcal D} \bigg) \partial_y \psi_n \, = \, - \, e^{-\frac{4 k y}{\mathcal{D}}} \, M_n^2 \psi_n \;,
\end{equation}
which is of the same form as the one considered in~\cite{Choi:2017ncj}. The normalized clockwork 0-mode, with Neumann boundary conditions, is easily found to be
\begin{equation}\label{eq:suppression}
\psi_0 \, = \, \mathcal{C} \, \equiv  \,\sqrt{\frac{D-3}{\mathcal{D}} \, \frac{\pi k R}{e^{\frac{D-3}{\mathcal{D}} 2 k \pi R}-1}} \;,
\end{equation}
which has the same form as the usual one~\cite{Giudice:2016yja}, with the replacement $k \to k (D-3)/\mathcal{D}$. Notice that $\mathcal{C}$ gives precisely the exponential clockwork suppression of the interactions between the clockwork 0-mode and a sector localized at $y = 0$ (e.g.~the Standard Model), relatively to the ones between the Kaluza-Klein clockwork gears and the same sector. 

\begin{figure}
\centering
\includegraphics[width=0.8\columnwidth]{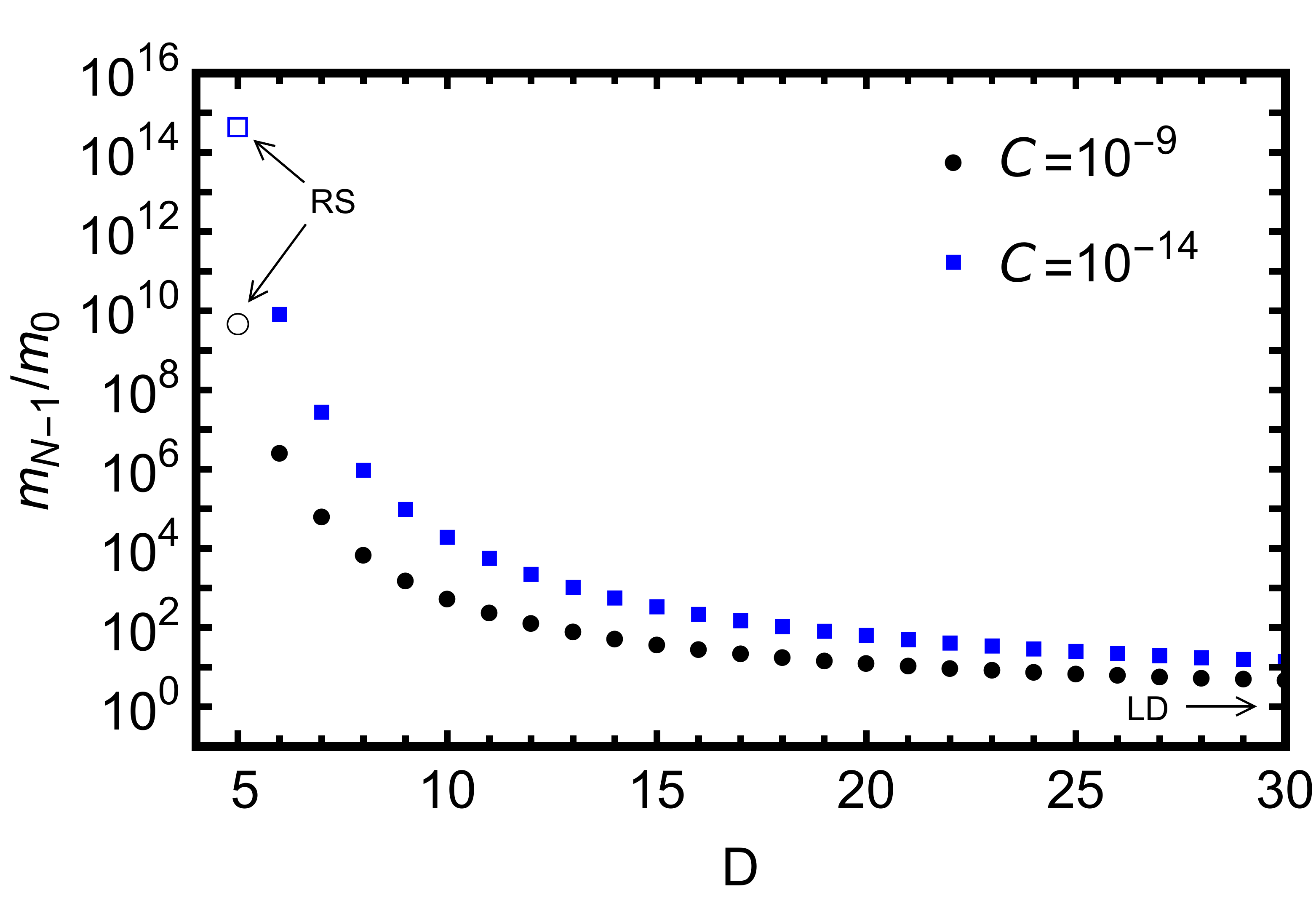}
\caption{Ratio of the parameters $m_{N-1}$ and $m_{0}$ in the discretized Lagrangian, for large $N$. For each value of $D$, the product $k R$ is chosen by means of~\eqref{eq:suppression} so that the clockwork suppression $\mathcal{C}$ is fixed. We also show the values obtained in the deconstructed Randall-Sundrum (at $D=5$) and in the linear dilaton (for $D \to \infty$) models.
\label{fig:hierarchy}}
\end{figure}

To show that for large $D$ this is indeed a clockwork model, let us discretize the extra dimension on $N$ points, obtaining the clockwork chain
\begin{equation}
- \frac{1}{2} \int \! d^4 x \sum_n \, \big[ (\partial_\mu \phi_n)^2  \, + \, m_n^2 (\phi_n - q \phi_{n+1})^2\big] \;,
\end{equation}
with
\begin{equation}
q = e^{\frac{D-3}{\mathcal{D}} \frac{k \pi R}{N} } \;, \quad m_n = \frac{N}{\pi R} \, e^{\frac{2 k \pi R} {\mathcal{D} N} n } \;.
\end{equation}
As shown in Fig.~\ref{fig:hierarchy}, although the Lagrangian parameters $m_n$ vary along the clockwork chain, in the limit of large $D$ there is no large hierarchy between them, and still an exponential clockwork suppression is generated. This is precisely the definition of the clockwork outlined in Sec.~\ref{sec:intro}, so that for large but finite $D$ we still have a clockwork theory. 

Fig.~\ref{fig:hierarchy} also shows that the quasi-clockwork considered in this section interpolates between the Randall-Sundrum and the linear dilaton models. As for the former, it is not surprising that this is recovered when the number of dimensions $D$ is analytically continued to $D = 5$: in this limit one does not have the additional extra dimensions of $\mathcal{F}_{D-5}$ at all, and the setup is precisely the Randall-Sundrum one, i.e.~a single orbifolded extra dimension with a negative bulk cosmological constant.

\begin{figure}
\centering
\includegraphics[width=0.8\columnwidth]{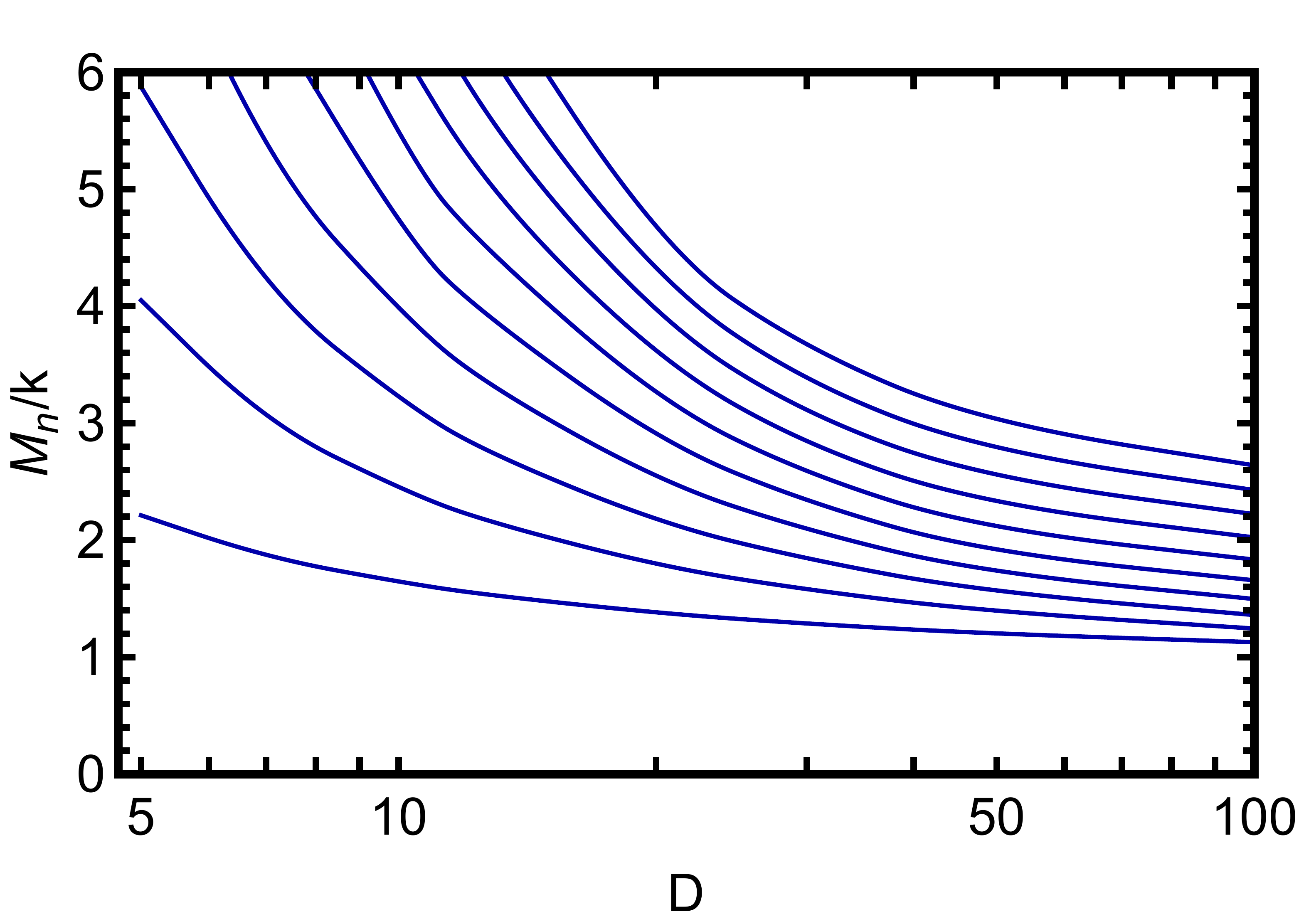}
\caption{Mass of the first ten Kaluza-Klein modes as a function of $D$. For each value of $D$, the product $k R$ is fixed by means of~\eqref{eq:suppression} so that the clockwork suppression is $\mathcal{C} = 10^{-6}$. \label{fig:gears}}
\end{figure}

Finally, the Kaluza-Klein spectrum is calculated in~\ref{app:gears} and shown in Fig.~\ref{fig:gears} as a function of $D$. Again, one interpolates between the Randall-Sundrum and the linear dilaton models. For large $D$ there is a mass gap $\simeq k$ and the gears are quasi-degenerate, whereas analytically continuing to $D=5$ one recovers the widely-spaced Randall-Sundrum spectrum. These results show that, although the precise form of the spectrum is model-dependent, the quasi degeneracy of the gears is a robust feature of clockwork models.

\section{Conclusions}

We have shown that the clockwork mechanism can be obtained, in a robust way, without requiring supersymmetry to forbid the fatal 5D cosmological-constant term that would otherwise invalidate the mechanism. Once again, we stress here that without a protection mechanism this would be a tuning \emph{in addition} to the one for the 4D cosmological constant, thus invalidating the clockwork solution to fine-tuning issues, such as the hierarchy problem.

 Quite remarkably, the construction presented here is rather minimal, showing that the clockwork setup is obtained rather generically from pure gravity in a large number of sufficiently flat  extra dimensions, without necessarily relying on string-theoretical constructions. The linear-dilaton EFT is obtained in the limit $D \to \infty$. Of course we need to assume, as in the linear-dilaton or Randall-Sundrum cases, that the brane terms are extremely non-generic, being tuned to support a 4D Minkowski spacetime (i.e.~the usual tuning of the cosmological constant mentioned above).

In addition, and as in the linear-dilaton or Randall-Sundrum cases, one needs a stabilization mechanism for the fifth ``clockwork'' dimension. If this is achieved by means of the dilaton itself~\cite{Cox:2012ee} one needs to generate non-analytic brane terms~\cite{Giudice:2017fmj}. Although it has been argued that this is in principle possible~\cite{Antoniadis:2001sw}, it would be interesting to study this issue explicitly. Otherwise, one could use the Goldberger-Wise mechanism~\cite{Goldberger:1999uk} by introducing a scalar field in the bulk.

While a robust clockwork setup is obtained for finite-$D$, the linear dilaton model is recovered only in an asymptotic sense, i.e.~one has a class of robust models that look arbitrarily close to it. However, it would be interesting to investigate if there are constructions, along the lines of this letter, that give the linear dilaton model for finite $D$ in a robust way.

Finally, let us quickly comment on the phenomenological predictions for the finite-$D$  construction, relatively to the more familiar linear-dilaton case, studied in detail in~\cite{Giudice:2017suc}. For $k \sim 100-1000$ GeV, one could hope to see the first gears as distinct resonances at the LHC. In this case, if $D$ is not too large, one could potentially distinguish the shape of the spectrum from the linear dilaton one (see Fig.~\ref{fig:gears}). The best prospects for the linear dilaton model are to look for  modifications  of the continuum $\gamma \gamma$ and $\ell \ell$ spectra, due to the closely-packed clockwork gears. In our construction, there are additional extra dimensions at a scale $L \ll R$. The continuum spectrum will be modified, with respect to the one calculated in~\cite{Giudice:2017suc}, starting at a scale $\sim 1/L$, even for $D \to \infty$. This is, in principle, an observable effect if the hierarchy between $L$ and $R$ is not too large. For instance, for $k = 300$ GeV, one has $R \approx 1/30$ GeV$^{-1}$ and the continuum spectrum gets distorted before, say, 1 TeV, for $L \gtrsim 0.03 R$. For $k = 10$ GeV the same is obtained for $L \gtrsim 10^{-3} R$.

\section*{Acknowledgements}
\vspace{-3mm}
\noindent
We thank Tony Gherghetta and Riccardo Argurio for comments. This work is supported by an ULB fellowship, an ULB-ARC grant and the IAP P7/37.

\appendix

\makeatletter
\gdef\thesection{\appendixname~\@Alph\c@section}%
\makeatother

\section{Consistency of the Kaluza-Klein ansatz} \label{app:KK_consistency}
Consistency of the Kaluza-Klein ansatz~\eqref{eq:metric} demands that the equations of motion as obtained from the $D$-dimensional and 5D theory are the same. This is known to be the case in the absence of a cosmological constant, as long as the ansatz contains the relevant degrees of freedom. It is worth to show explicitly that this is still the case in the presence of $\widetilde \Lambda$.

Let us start from the $D$-dimensional theory. In this case the equations of motion are the Einstein field equations in the presence of a cosmological constant $\widetilde \Lambda$. For the metric~\eqref{eq:metric_general} one finds
\begin{align}
\widetilde{\mathcal{R}}_{mn} &= \, \mathcal{R}_{mn}- \nabla_m \partial_n S - \frac{1}{D-5} \,\partial_m S \, \partial_n S\;,\\
\widetilde{\mathcal{R}}_{ab} \ &= - \, \gamma_{ab} \, \frac{e^{\frac{2 S}{D-5}}}{D-5} \Big( \nabla^2 S \,+\, \partial^m S \, \partial_m S\Big) \;,\\
\widetilde{\mathcal{R}} \ &= \, \mathcal{R} - \, 2  \, \nabla^2 S \, - \, \frac{D-4}{D-5} \, \partial^m S \, \partial_m S \;,
\end{align}
where the covariant derivatives are with respect to the 5D part of the metric $g_{mn}$. However, it is convenient, in order to compare with the dimensionally-reduced theory in the Einstein frame, to perform a Weyl transformation $\widetilde{g}_{MN} \to e^{-\frac{2 S}{D-2}}\widetilde{g}_{MN}$. One finds 
\begin{align}
\widetilde{\mathcal{R}}_{mn} &\to \widetilde{\mathcal{R}}_{mn} + \nabla_m \partial_n S + \frac{g_{mn}}{D-2} \nabla^2 S + \frac{1}{D-2} \partial_m S \, \partial_n S \;,\\
\widetilde{\mathcal{R}}_{ab} &\to \widetilde{\mathcal{R}}_{ab} + \gamma_{ab} \, e^{\frac{2 S}{D-5}} \bigg( \frac{1}{D-5} \partial^m S \, \partial_m S + \frac{1}{D-2} \nabla^2 S\bigg) \, ,\\
\widetilde{\mathcal{R}} &\to e^\frac{2 S}{D-2} \bigg( \widetilde{\mathcal{R}} \,+\, 2 \, \frac{D-1}{D-2} \, \nabla^2 S  \,+\, \frac{D-1}{D-2} \, \partial^m S \, \partial_m S \bigg) \;,
\end{align}
having taken into account that, for instance, $\widetilde \nabla^M \partial_M S = \nabla^m \partial_m S + \partial^m S \partial_m S$. From these results, the Einstein field equation
\begin{equation}
\widetilde{\mathcal{R}}_{MN} \, - \, \frac{1}{2} \, \widetilde{g}_{MN} \widetilde{\mathcal{R}} \, + \, \widetilde{\Lambda} \, \widetilde{g}_{MN} \ = \ 0 \;, 
\end{equation}
takes on the form
\begin{align}
&\mathcal{G}_{mn} \ = \ \frac{3 \, ( \partial_m S \, \partial_n S - \frac{g_{mn}}{2} \partial^r S \, \partial_r S )}{(D-2)(D-5)} \,-\, \widetilde{\Lambda} \, g_{mn} e^{-\frac{2 S}{D-2}}\;,\label{eq:EoM1}\\
&0 = - \frac{\mathcal{R}}{2} - \frac{\nabla^2 S}{D-5} + \frac{3 \, \partial^m S \, \partial_m S}{2\, (D-2)(D-5)} \, + \, \widetilde{\Lambda} \, e^{- \frac{2S}{D-2}} \;,\label{eq:EoM2}
\end{align}
with $\mathcal{G}_{mn} \equiv \mathcal{R}_{mn} - g_{mn} \mathcal{R}/2$.

Let us now move to the dimensionally-reduced theory. The 5D Jordan-frame action is~\eqref{eq:bulk_Lagr} which, by the field redefinition $g_{mn} \to e^{- \frac{2S}{3}} g_{mn}$, is turned into the Einstein-frame one:
\begin{equation}
\sqrt{-g} \mathcal{L} = \frac{M_5^3}{2}   \sqrt{-g} \, \bigg( \mathcal{R} \,- \frac{1}{3}\,\frac{D-2}{D-5} \partial_m S \, \partial^m S \,-\, 2 \widetilde{\Lambda} \, e^{-\frac{2 S}{3}} \bigg) \;,
\end{equation}
from which one can obtain the 5D Einstein field equation
\begin{equation}\label{eq:EoM3}
\mathcal{G}_{mn} = \frac{1}{3} \frac{D-2}{D-5} \Big( \partial_m S \, \partial_n S - \frac{g_{mn}}{2} \partial^r S \partial_r S \Big)  -  \widetilde{\Lambda} g_{mn} e^{-\frac{2}{3}S} \;.
\end{equation}
This coincides with~\eqref{eq:EoM1} after the field redefinition $S/3 \to S/(D-2)$. The equation of motion for $S$ is
\begin{equation}
\frac{D-2}{D-5} \, \nabla^2 S \,+\, 2 \widetilde{\Lambda} \, e^{- \frac{2}{3} S} = 0\;,
\end{equation}
which coincides with~\eqref{eq:EoM2} after combining with the trace of~\eqref{eq:EoM3} and redefining $S/3 \to S/(D-2)$. Therefore, the equations of motion are the same in the $D$-dimensional theory and in the dimensionally-reduced 5D one.

\section{Curved compactifications}\label{app:curved}
In this appendix we study what happens when the $D-5$ extra dimensions are not flat. For simplicity, let us consider the case of constant Ricci scalar, $\mathcal{R}(\gamma_{ab}) = \mathcal{K}$. In the case of a $D-5$ dimensional sphere, i.e.~if the flat manifold $\mathcal{F}_{D-5}$ is replaced by $\mathcal{S}_{D-5}$, we have $\mathcal{K} = (D-5)(D-6)/r^2$, where $r$ is the coordinate radius, which requires some unspecified stabilization. Pure gravity in $D$ dimensions, without the cosmological constant $\widetilde{\Lambda}$, is now dimensionally reduced to
\begin{align}
\frac{M_5^3}{2}  \int_{\mathcal{M}_{CW}} \hspace{-1.5em} d^5 x \, \sqrt{-g} \, e^S \Big( \mathcal{R} \,+\, \frac{D-6}{D-5} \, \partial_m S \, \partial^m S \,+\, \mathcal{K} \, e^{- \frac{2 S}{D-5}} \Big) \;.
\end{align}
For large $D$ the effect of the curvature mimics the effect of $k$ in~\eqref{eq:bulk_Lagr}. The linear-dilaton EFT is obtained in the limit $D\to \infty$. However, one also needs $\mathcal{K}$ to stay finite in this limit. This implies that the $D-5$ extra dimensions become asymptotically flat, as can be seen from the case of a sphere: the curvature radius is $r \, e^{-\frac{S}{D-5}} \approx r \to \infty$, to have finite $\mathcal{K}$. Nevertheless, the case of a sphere is not quite successful: for finite $D$ and in the presence of $k$ as in~\eqref{eq:bulk_Lagr}, one needs to have $r \gtrsim (D-5) R/20$, for $k R \approx 10$, and at the same time $r \ll R$ in order to be able to dimensionally reduce the theory to 5D.

More in general, one may be worried that the curvature needs to be tuned to be exponentially small, with a tuning analogous to~\eqref{eq:tuning}. We now show that this is not the case. Let us consider a small curvature $\mathcal K = 2 \epsilon k^2$, with $\epsilon \ll 1$. The potential in the Einstein frame is
\begin{equation}\label{eq:V}
V = \frac{M_5^3}{2} \Big( - 4 k^2 e^{- \frac{2 S}{3}} - 2 \epsilon k^2 e^{- \frac{2}{3} \frac{D-2}{D-5} S} \Big) \;.
\end{equation}
The solution to the bulk equations of motion can be found approximately in the limit $\epsilon \ll 1$ by using the ``superpotential'' technique~\cite{DeWolfe:1999cp}. To this purpose, it is convenient to work in warped coordinates: $d s^2 = e^{2 \sigma(z)} \eta_{\mu \nu} dx^\mu dx^\nu + dz^2$. In this system of coordinates, the unperturbed solution with $\epsilon = 0$ is:
\begin{equation}\label{eq:unpert}
  S(z) = 3 \ln \frac{z}{z_0} \;,  \qquad \sigma(z) = \frac{D-6}{D-5} \,  \ln \frac{z}{z_0} \;,
\end{equation}
with $2 k z_0 \equiv \sqrt{(3D - 18)(3D - 19)}/(D-5)$. The perturbed potential~\eqref{eq:V} is obtained from a ``superpotential'' of the form $W= A e^{B \Phi} + \eta e^{C \Phi}$, where $\Phi$ is the canonically normalized field $S$ and $\eta$ is $O(\epsilon)$. One finds that the solution~\eqref{eq:unpert} is perturbed by
\begin{align}
\delta S(z) \ &=\, \frac{3 \epsilon}{4} \, \bigg(\frac{z_0}{z}\bigg)^{\!\!\frac{6}{D-5}} \, \frac{(D+1)(3 D - 19)}{(D-11)(3D-25)} \;,\\
\delta \sigma(z) \ &=\, - \frac{\epsilon}{2} \, \bigg(\frac{z_0}{z}\bigg)^{\!\!\frac{6}{D-5}} \, \frac{(D-6)(3 D - 19)}{(D-11)(3D-25)} \;,
\end{align}
for $D \neq 11$. Therefore, the curvature (as measured by the Ricci scalar) does not need to be exponentially small, since $z \geq z_0$ and $D>5$. This result could have also been anticipated by noting that the exponential term in~\eqref{eq:V} related to the curvature does vanish faster than the usual $k$-term, contrary to what would happen for a bulk cosmological constant.

\section{Quasi-clockwork gears}\label{app:gears}
In this appendix we find the Kaluza-Klein spectrum for finite $D$. To solve the equation of motion~\eqref{eq:EoM} it is convenient to introduce $f_n = \psi_n e^{\frac{D-3}{\mathcal{D}}k y}$, for which one has
\begin{equation}
\bigg( \partial_y^2 + \frac{4 k }{\mathcal D} \partial_y - M_f^2 \bigg) f_n = - M_n^2 \, e^{- \frac{4 k y}{\mathcal D}} f_n \;,
\end{equation}
with $M_f^2 = k^2 (D+1)(D-3)/\mathcal{D}^2$. This differential equation is familiar from Randall-Sundrum models~\cite{Gherghetta:2000qt}. One obtains:
\begin{align}
&\psi_n = \frac{1}{\mathcal{N}_n} \, e^{- \frac{D-1}{\mathcal D} k y} \, \times \notag\\
&\bigg[ J_{\frac{D-1}{2}}\!\Big( \frac{\mathcal{D} M_n}{2 k} \, e^{-\frac{2 k }{\mathcal{D}} y} \Big) + b_n(M_n) \, Y_{\frac{D-1}{2}}\!\Big( \frac{\mathcal{D} M_n}{2 k} \, e^{-\frac{2 k }{\mathcal{D}} y} \Big) \bigg] \;,
\end{align}
where the Neumann boundary condition at $y=0$ gives
\begin{equation}
b_n(M_n) \equiv - \, \frac{J_{\frac{D-3}{2}}\!\Big( \frac{\mathcal{D} M_n}{2 k}\Big)}{Y_{\frac{D-3}{2}}\!\Big( \frac{\mathcal{D} M_n}{2 k}\Big)} \;,
\end{equation}
and the one at $y=\pi R$,
\begin{equation}
b_n(M_n) = b_n(M_n \, e^{-\frac{2 \pi k R}{\mathcal D}}) \;,
\end{equation}
allows to find numerically the spectrum plotted in Fig.~\ref{fig:gears}.


\bibliography{clockwork}

\end{document}